\documentclass[aps,preprint,showpacs,showkeys]{revtex4}
\usepackage{amsfonts}
\usepackage{subfigure}
\usepackage{amsmath}
\usepackage{amssymb}
\usepackage{graphicx}
\usepackage{epsfig}

\begin{document}

\preprint{}
\title[ ]{Zitterbewegung, internal momentum and spin of the circular
travelling wave electromagnetic electron}
\author{Malik Mohammad Asif}
\email{masif@comsats.edu.pk}
\author{Salman Khan}
\email{sksafi@comsats.edu.pk}
\affiliation{COMSATS Institute of Information Technology, Park Road, Tarlai Kalan,
Islamabad 45550, Pakistan}
\date{December 14, 2015}

\begin{abstract}
The study of this paper demonstrates that electron has Dirac delta like
internal momentum ($u,\overrightarrow{p}_{\theta }$), going round in a circle of
radius equal to half the reduced Compton wavelength of electron with
tangential velocity $c$. The circular momentum $\overrightarrow{p}_{\theta }$ and
energy $u$ emanate from circular Dirac delta type rotating monochromatic $\ $%
electromagnetic (EM) wave that itself travels in another circle having
radius equal to the reduced Compton wavelength of electron. The phenomenon
of Zitterbewegung and the spin of electron are the natural consequences of
the model. The spin is associated with the internal circulating momentum of
electron in terms of four component spinor, which leads to the Dirac
equation linking the EM electron model with quantum mechanical theory. Our
model accurately explains the experimental results of electron channelling
experiment, [P. Catillon et al., Found.Phys. $38$, $659$ ($2008$)], in which
the momentum resonance is observed at $161.784MeV/c$ corresponding to
Zitterbewegung frequency of $80.874MeV/c$ electron beam.
\end{abstract}

\maketitle

\section{Introduction}

The electron is elementary particle, embodied with intrinsic properties such
as charge, mass, spin and magnetic dipole moment. Historically, there are
number of proposed extended classical models of electron \cite{Jansen,
Barut, Dirac2, Dirac3, Burinskii, Qiu Hong} encompassing classical
properties effectively. However, these models not only suffer from the
electron instability problem but also are unable to explain the quantum
mechanical (QM) behavior of electron. On the other hand, the celebrated QM
model of Dirac \cite{Dirac 1} is an abstract mathematical model, which
incorporates the intrinsic properties of electron in terms of Dirac matrices
and $4$-component wavefunction. Dirac's later works on extended model of
electron \cite{Dirac2, Dirac3, Dirac 4} attempts to expound its classical
properties. However, on the conceptual background the comprehension of the
abstract point QM model and its possible association to electrons's
properties is still not well clear.

While working on Dirac electron in an external electromagnetic field,
Foldy-Wouthuysen noted that the particle spreads out over a region of
dimension of the order of its Compton wavelength in space with its mean
position $X^{^{\prime }}=x$. This behavior of the particle lead to the
proposal of extended model of electron \cite{Foldy L. L, Franz Schwabl}. Due
to some reasons, the extended models or their interpretation such as the one
given in \cite{Foldy L. L} didn't get much attention. The detail studies on
Dirac model by Schr\"{o}dinger and others \cite{E. Schrodinger, Barut and A.
J} show that the electron undergoes an internal oscillatory motion having
dimension of the order of Compton wavelength, called Zitterbewegung (Ztg).
From the Dirac equation of motion for a free electron, Schr\"{o}dinger
showed that even in linear motion the electron executes an oscillatory
motion with a frequency $\nu _{o}=2m_{o}c/h$, where all the parameters bear
their traditional meaning. He further concluded that the instantaneous
velocity of a slowly moving electron is $c$.

Recently the electron model proposed by Alexander Burinskii \cite{Burinskii,
Burinskii 1, Burinskii 2} considers electron as a closed gravitational
singular string, based on Kerr-Newman theory. The model is essentially in
curved space time geometry. Similarly, the electron model \cite{Atiyah} uses
($4+1$) Ricci flat space time geometry to obtain exact solutions of the
Maxwell's and gauged Dirac equations. These solutions are interpreted in
terms of a geometric model of the electron with its spin. Hence these models
are inherently geometric ones, based on curved or ($4+1$) space-time,
instead of a flat spacetime geometry. Again these models do not institute
some physical mechanism(s) at the core level. Using the Maxwell's theory
with the boundary condition of spherical conducting surface, Dirac presented
an electron model \cite{Dirac 4}, which explains the existence of muon as
the excited state of electron, however, it does not assign spin to the
electron.

In a very recent study, a traveling wave electromagnetic (TWEM) model for
electron is presented \cite{M.M.Asif}. The model is based on rotating EM
wave confined in a circle with radius equal to the reduced Compton
wavelength (RCW) ($\hslash /mc$) of electron using Maxwell's theory as a
physical mechanism at the core level. In TWEM model the electric field
vector $\vec{E}_{r}$ (a Dirac delta vector) rotates in the $xy$ plane and
the corresponding magnetic field vector $\vec{H}_{z}$ is pseudo static Dirac
delta vector along the $z$-direction. The EM wave traverses a circular
trajectory with RCW of the electron in half space. The radius and energy of
the modeled electron, matches precisely to the classical counterparts. The
stability of EM electron is ensured by showing zero divergence of
source-free EM\ energy-momentum stress tensor for the model. The charge
generation mechanism has been discussed at length using gauge of EM\ fields.
The time reversal in travelling wave corresponds to the reversal of $\vec{E}$
and $\vec{B}$ fields, which in turn gives rise to antiparticle (the
positron) rotating EM wave. With four possible time-space ($\pm t,\pm \theta
$) combinations we obtain four types of EM rotating waves. The overall
picture at this level conforms the local $U(1)$ gauge symmetry. The
monochromatic circulating EM\ wave, carry the corresponding energy-momentum
in half the RCW of electron. The Ztg phenomenon arises naturally in this
model as a consequence of circular flow of EM energy- momentum.

In this paper we use the TWEM model to demonstrate the electron's internal
structure in terms of circular flow of energy-momentum ($u_{em,}\ \vec{p}%
_{\theta }$) at velocity $c$. The two component wavefunction for
energy-momentum, corresponding to left and right circular polarized wave(s)
is shown to be related to the spin of electron, conforming to $SU(2)$ group
symmetry. Furthermore, the negative and positive time circulating
energy-momentum are shown to correspond to four component Dirac like spinor.
In the rest frame, the four components of spinor satisfy massless Klein
Gordon equation, which in turn leads to a massless Dirac or Weyl equation,
governed by transformation rule under covering group $SL(2,C)$ of proper
Lorentz group $SO^{+}(3,1)$. Finally, the massive Dirac equation for TWEM
electron has been obtained in boosted frame linking EM electron model with
QM.

The internal dynamical structure of electron based on Ztg can be
experimentally verified in line with experiments of electron channeling in
silicon crystals \cite{Catillon, Gouan}. In electron channeling experiment,
the momentum resonance is observed at twice the de Broglie frequency, that
is at $161.784MeV/c$ instead of $80.874MeV/c$. The authors describe it as a
manifestation of de Broglie internal clock if squared amplitude is
considered, that is, the Ztg frequency. In our proposed model the frequency
of momentum is twice the de Broglie frequency of EM circulating wave. This
is in good agreement with the phenomena observed in the electron channelling
experiment.

\section{Zitterbewegung and the velocity of electron}

In this section, for the paper to be self-contained, we present a brief
review of the basic concepts of Ztg in electron. Although a number of
different interpretations of Ztg of Dirac model are given \cite{Foldy L.
L,David Hestenes 1,David Hestenes 2,Sasabe, E. Romera, G. Cavalleri, David
Hestenes 3, D. Hestenes, K. Huang, Zhi-Yong}, however, we will briefly
review the interpretation given in \cite{E. Schrodinger, Barut and A. J}.
The Dirac equation for spin-$1/2$ particle is%
\begin{equation}
i\hbar \frac{\partial }{\partial t}\psi (r,t)=H\psi (r,t),  \label{1}
\end{equation}%
where $H$ is the Hamiltonian for a free particle and can be written as
\begin{equation}
H=c\vec{\alpha}\cdot \vec{p}+\beta mc^{2}.  \label{2}
\end{equation}%
In Eq. (\ref{2}) $\vec{\alpha}$ and $\beta $ are the well known $4\times 4$
Dirac matrices which satisfy following conditions,%
\begin{equation}
\beta ^{2}=I;\quad \alpha _{i}\alpha _{j}+\alpha _{j}\alpha _{i}=2\delta
_{ij}I;\quad \alpha _{i}\beta +\beta \alpha _{i}=0.  \label{2a}
\end{equation}%
For a free electron, the velocity determined by Schr\"{o}dinger \cite{E.
Schrodinger}, using Eq. (\ref{2}) is
\begin{equation*}
\vec{v}=c\vec{\alpha},
\end{equation*}%
with $c$ being the expectation value of $\overrightarrow{v}$, which means the electron is moving with an aggregate constant velocity $c$.
It was further illustrated that $\vec{v}$ can be expressed as.
\begin{equation}
\vec{v}=c^{2}H^{-1}\vec{p}\mathbf{+}c\vec{\eta}_{o}e^{-2iHt/\hbar },
\label{3}
\end{equation}%
where $\vec{p}$ is the linear momentum and $\vec{\eta}_{o}$ is a constant
operator given by
\begin{equation*}
\vec{\eta}_{o}=\vec{\eta}_{o}\mathbf{(}0\mathbf{)=}\vec{\alpha}\mathbf{(}0%
\mathbf{)-}cH^{-1}\vec{p}\mathbf{.}
\end{equation*}%
The velocity of electron in Eq. (\ref{3}) consists of two parts, the regular
part $c^{2}H^{-1}\vec{p}$ and the oscillatory part $c\vec{\eta}%
_{o}e^{-2iHt/\hbar }$. The later part represents Ztg having amplitude equal
to $\hbar /2m_{o}c$ (half the RCW of electron) and angular frequency $\omega
_{z}=2m_{o}c^{2}/\hbar $. The oscillatory part, having extremely small
amplitude, was believed unobservable directly \cite{P.A.M. Dirac 1}.
However, a recent simulated work \cite{R. Gerritsma} on one dimensional
Dirac dynamics for free particle shows that the Ztg is an observable
phenomenon. Also, \cite{J. Y. Vaishnav} proposes an experiment with
ultracold atoms in a tripod level scheme on an optical lattice, to observe
Ztg, at experimentally measurable frequencies. The \textit{internal dynamics
of 4 momentum} associated with the oscillatory part of electron is the
source of Ztg of electron and is at the core of QM\ theory embedded in $%
\alpha _{i}$ and $\beta $ of Dirac matrices.

\section{Internal 4-momentum of Circular EM travelling wave electron}

The relativistic momentum $\vec{p}^{\ \prime }$ of a free electron is given
as%
\begin{equation}
\left\vert \vec{p}\ ^{\prime }\right\vert =\pm \sqrt{\vec{p}^{\ 2}+\vec{p}%
_{o}^{\ 2}},  \label{4}
\end{equation}%
where $\vec{p}$ is the linear momentum in $x,y,z$ coordinates and $%
p_{o}=m_{o}c$ is the magnitude of momentum associated with the rest mass of
electron. The momentum $p_{o}$ is Lorentz invariant and behaves as a
constant in the frame of the center of mass of the particle. However, there
is dynamism of energy-momentum when electron is at rest, as noted by
Hestenes \cite{David Hestenes 3} and others \cite{H. Torres-Silva}. In this
article we will consider the internal momentum dynamism (spin) in the center
of mass frame of electron, which naturally arises in the TWEM model of
electron \cite{M.M.Asif}. The $EM$\ momentum of electron in the particle's
rest frame $K_{p}(0)$ is given by%
\begin{equation}
\vec{p}_{o}\equiv \vec{p}_{\theta }=\frac{1}{c}\vec{S}_{\theta }.  \label{5}
\end{equation}%
The angular momentum $\vec{p}_{\theta }$ is invariant momentum associated
with the spin of electron. In the rest frame, we have $\vec{p}=0$ in Eq. (%
\ref{4}) and $p^{\prime }=\left\vert \pm \vec{p}_{\theta }\right\vert =$ $%
\left\vert \frac{1}{c}\vec{S}_{\theta }\right\vert $, which leads to the
massless -light like case.

Using the internal momentum $\vec{p}_{\theta }$ and energy $%
u_{em}=(\varepsilon _{o}\vec{E}_{r}^{2}+\mu _{o}\vec{H}_{z}^{2})/2$ of
electron in the rest state, we can define internal momentum of electron as
follows%
\begin{equation}
p_{\theta }^{\mu }=(p_{\theta }^{o},\vec{p}_{\theta }),  \label{6a}
\end{equation}%
where $p_{\theta }^{o}=u_{em}/c$. The $p_{_{\theta }}^{\mu }$ apparently
forms $4$-vector with $p_{\theta }^{o}$, and $\vec{p}_{\theta }$ the time
like and spatial component(s), respectively, in the rest frame of electron.
It, however, does not transform as $4$-vector, since $p_{\theta }^{\mu }$ is
the internal degree of freedom of electron associated with the spin and
subsist as a $2D$ object. In the rest frame of electron, $p_{\theta }^{\mu }$
can be considered to dwell in wrapped coordinates within the particle \cite%
{Barut and A. J, GIOVANNI}.

\begin{figure}[h]
\begin{center}
\includegraphics[scale=0.9]{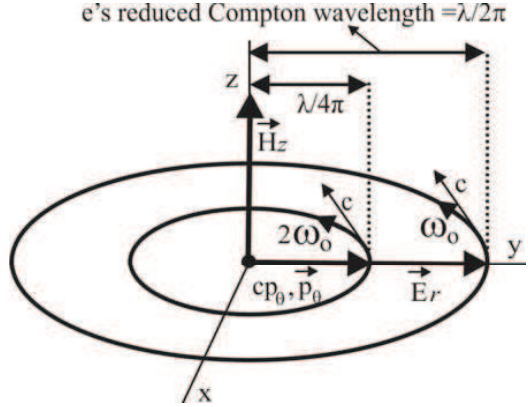}
\end{center}
\caption{The vector $\vec{E}_{r}$ is circulating in a circle of redius equal
to RCW and the corresponding momentum $\vec{p}_{\protect\theta }$ and energy
$cp_{\protect\theta }$ are circulating in a circle of radius equal to $\frac{%
1}{2}$RCW. The vectors $\vec{E}_{r}$ and $\vec{p}_{\protect\theta }$ have
angular frequency $\protect\omega _{o}$ and $2\protect\omega _{o}$,
respectively, but the same tangential velocity $c$.}
\end{figure}

Using the circular EM\ travelling wave expressions for TWEM\ model of
electron, the $\mathbf{p}_{\theta }$ can be expressed as,%
\begin{equation}
\vec{p}_{\theta }=\frac{1}{c}p_{o}^{^{\prime }}e^{\pm i/\hslash
(-2u_{em}t\pm 2p_{\theta }r\theta )}\hat{\theta},  \label{7}
\end{equation}%
where $p_{o}^{\prime }=\frac{1}{c}[E_{r}^{o}H_{z}^{o}]$. We have a similar
expression for the energy (time like) component $p_{\theta }^{o}=cp_{\theta
} $. There are four EM\ rotating (circular polarized like) waves with
circulating momentum given in Eq. (\ref{7}). Two of these are associated
with electron in positive ($+ve$) \textit{time} (or $+ve$ energy), right ($%
\uparrow $)$\equiv \psi ^{1}$ and left circular ($\downarrow $) $\equiv \psi
^{2}$ and other two associated with positron in negative ($-ve)$ \textit{time%
} (or $-ve$ energy), right ($\uparrow $) $\equiv \psi ^{3}$ and left
circular ($\downarrow $) $\equiv \psi ^{4}$. Therefore, we can define four
spinors $\psi ^{i}$ ($i=1,2,3,4$) corresponding to four circulating
polarized waves with four time-space ($\pm t,\pm \theta $) combinations.

Initially, we consider the case of $\psi ^{1}$ (in $+t,+\theta $
coordinates) and will extend the idea at the end to encompass four component
$\vec{\psi}\cong \psi ^{\alpha };$ $\alpha =1,2,3,4$, to arrive at Dirac
equation. The momentum in coordinate combination ($+t,+\theta $ ) is%
\begin{equation}
\vec{p}_{\theta }=p_{o}^{\prime }[e^{-i(\omega _{o}^{\prime }t-k_{\theta
}^{\prime }\mathbf{\theta })}]\hat{\theta},  \label{8}
\end{equation}%
with $\omega _{o}^{\prime }=2\omega _{o}$ and $k_{\theta }^{\prime }=$ $%
2\omega _{o}r/c$. In this case the electric field vector $\vec{E}$ ($%
+t,+\theta $ ) rotates with angular velocity $\omega _{o}$, in a circle of
radius $r_{o}\equiv $RCW. In Eq. (\ref{8}), we note that the momentum $\vec{p%
}_{\theta }$ rotates with twice the angular velocity of $\vec{E}_{r}$
vector, in a circle of radius $r_{o}/2$. From Eq. (\ref{8}), we determine
the angular velocity (or frequency) of energy-momentum as follows%
\begin{equation}
\omega _{\theta }^{p}=\omega _{\theta }^{u}=\frac{\omega _{o}^{\prime }}{%
k_{\theta }^{\prime }}=\frac{c}{r_{o}},  \label{9}
\end{equation}%
where $\omega _{\theta }^{p}$ and $\omega _{\theta }^{u}$ are the angular
velocities of momentum and energy, respectively. The corresponding
tangential velocities are $\left\vert \vec{v}^{p}\right\vert =\left\vert
\vec{v}^{u}\right\vert =c$, (see Fig. (\ref{1})). The angular frequency $%
\omega _{o}^{\prime }$ of energy-momentum is in conformity with Ztg angular
frequency \cite{E. Schrodinger, Barut and A. J}.

As both the momentum $\vec{p}_{\theta }$ and the energy $p_{\theta }^{o}c$
are circulating with the velocity of light, hence, we will first consider
massless Dirac or Weyl spinor within $(1/2)$ RCW of electron. In fact, the
Schr\"{o}dinger result for the expectation value of $\overrightarrow{v}$ equals $c$ for Ztg of
electron is valid only if the electron (or its constituent spinors) behave
as massless object(s) in the relevant region \cite{Qiu Hong}. We will
however see that with boost, the left and the right spinors combine through
the mass term. This leads to the actual velocity of the particle in the
moving frame and the spinors still circulate at $c$, the velocity of light.

The extended TWEM model of electron\ exhibits internal geometry in the form
of rotating transverse EM wave in a circle of radius $r_{o}$\ (RCW). That is
the source free, Dirac delta type EM\ fields exist in a sphere of radius $%
r_{o}$. The energy-momentum of EM\ travelling wave flows in a circle of
radius $(1/2)r_{o}$\ synchronously with EM\ wave in the $xy$\ plane with
speed $c$, as shown in Fig.(\ref{1}). The circulating energy-momentum
subsist in $SU(2)$\ symmetry group, bestows QM spin satisfying all $SU(2)$\
commutation relations. As we show below, the magnitude of $z$-component of
spin turns out to be $\hslash /2$\ and the helicity corresponding to right
circular or left circular states determine its spin up or down state,
respectively. A similar geometry of extended model of electron based on Ztg,
is described in \cite{Bruno}.

\section{Electron in the rest state}

For electron in the rest state, Eq. (\ref{8}) may be written as
\begin{equation}
\vec{p}_{\theta }=p_{\theta }e^{i\theta ^{\prime }}\hat{\theta},  \label{11}
\end{equation}%
where $\theta ^{\prime }=(2m_{o}r/\hslash )\theta $ and time exponential
part is implicit. Using cylindrical coordinates, we can write%
\begin{equation}
\vec{p}_{\theta }\equiv p_{x}\pm ip_{y}  \label{12}
\end{equation}%
This is the relation for momentum (and the corresponding energy), rotating,
counterclockwise (ccw) and clockwise (cw) in $xy$ plane, for $+$ and $-$
signs in Eq. (\ref{12}), respectively. Then using the analogy of circular
polarized waves as the combination of two orthogonal linear polarized waves
with a phase difference of $\pm \pi /2$, we get%
\begin{equation}
p_{\theta }e^{i\theta ^{\prime }}=p_{o}(\hat{x}\pm i\hat{y})e^{i\theta
^{\prime }}.  \label{12a}
\end{equation}%
Similar expression can be obtained for the energy $u_{em}$. The
energy-momentum taken together form four vector like object $p_{\theta
}^{\mu }$. Now, we can write two component wave function in momentum space $%
\psi ^{\nu }(\vec{p})$, for $\nu =1,2,$%
\begin{equation}
\psi (\vec{p})=\left[
\begin{array}{c}
\xi _{R} \\
\xi _{L}%
\end{array}%
\right] e^{i\theta ^{\prime }}=\left[
\begin{array}{c}
\psi ^{1} \\
\psi ^{2}%
\end{array}%
\right] ,  \label{12d}
\end{equation}%
where the right and the left eigenfunctions are, respectively, given by $\xi
_{R}\equiv (\xi _{R}^{u},\xi _{R}^{p})$ and $\xi _{L}\equiv (\xi
_{L}^{u},\xi _{L}^{p})$. The superscripts $u$ and $p$ denote the energy and
momentum parts, respectively. Moreover, the $4$-component spinor $\psi
^{\alpha }(\vec{p})$; $\alpha =1,2,3,4$ is obtained by including the $-ve$
time (or $-ve$ energy).

The four component wavefunction is the solution of the following second
order wave equation in cylindrical coordinates \cite{M.M.Asif},
\begin{equation}
\left[ \frac{1}{c^{2}}\frac{\partial ^{2}}{\partial t^{2}}-\frac{1}{r^{2}}%
\frac{\partial ^{2}}{\partial \theta ^{2}}\right] \psi ^{\alpha }(\vec{p})=0.
\label{12f}
\end{equation}%
This equation is analogous to massless Klein-Gordon equation in momentum
space. Using the standard procedure for obtaining first order differential
equation from second order, we get the massless Dirac equation for the
spinor $\psi ^{\alpha }(\vec{p})$%
\begin{equation*}
i\hslash \left[ \gamma ^{o}\frac{\partial }{\partial t}+\gamma ^{i}\frac{1}{r%
}\frac{\partial }{\partial \theta }\right] \psi ^{\alpha }(\vec{p})=0;\quad
i=1,2,3.
\end{equation*}%
Writing $\gamma ^{\mu }$ in Weyl representation

\begin{equation*}
\gamma ^{o}=\left[
\begin{array}{cc}
0 & I \\
I & 0%
\end{array}%
\right] ,\quad \gamma =\left[
\begin{array}{cc}
0 & \vec{\sigma} \\
\mathbf{-}\vec{\sigma} & 0%
\end{array}%
\right] ,
\end{equation*}%
where $I$ is $2\times 2$ identity matrix and $\vec{\sigma}$ are the Pauli
spin matrices. Using the units where $\hslash =c=1$, we arrive at Weyl
equations or massless Dirac equation in chiral representation%
\begin{equation}
\left[
\begin{array}{c}
(E-\vec{p}_{\theta }\cdot \vec{\sigma}\mathbf{)} \\
(E+\vec{p}_{\theta }\cdot \vec{\sigma}\mathbf{)}%
\end{array}%
\right] \left[
\begin{array}{c}
\vec{\psi}_{R} \\
\vec{\psi}_{L}%
\end{array}%
\right] =\left[
\begin{array}{c}
0 \\
0%
\end{array}%
\right] \mathbf{.}  \label{13a}
\end{equation}%
Here wave factor $e^{i\theta ^{\prime }}$ is implicit and $\vec{\psi}_{R}$
and $\vec{\psi}_{L}$ are $\times 2$ block wavefunctions

\begin{equation*}
\vec{\psi}_{R}\equiv \left[
\begin{array}{c}
\psi ^{1} \\
\psi ^{2}%
\end{array}%
\right] ,\quad \text{and}\quad \vec{\psi}_{L}\equiv \left[
\begin{array}{c}
\psi ^{3} \\
\psi ^{4}%
\end{array}%
\right] .
\end{equation*}%
We notice that $\vec{\psi}_{R}$ and $\vec{\psi}_{L}$ are independent and
there is no interference from one to the other in the rest state of electron.

Now, we define helicity operator for our case as
\begin{equation}
\hat{h}_{\theta }=\frac{\vec{\sigma}\cdot \vec{p}_{\theta }}{\left\vert \vec{%
p}_{\theta }\right\vert }.  \label{14}
\end{equation}%
The Weyl spinors are eigenstates of\ helicity operator $h=$ $\vec{\sigma}%
\mathbf{\cdot }\hat{n}$, with $\vec{p}/\left\vert \vec{p}\right\vert =\hat{n}
$. The helicity is well defined for the massless Dirac fermion or Weyl
fermion with eigenvalues equal to $\mathbf{\pm }1$. In this case we are
considering only internal momentum $\vec{p}_{\theta }$ ($2D$ object in the $%
xy$ plane), therefore, eigenvalues of $h_{\theta }$ are $+1$ and $-1$ for
ccw and cw rotation, respectively. Dirac massless case is also equivalent to
ultra relativistic helicity (energy $E>>m$) of particle (moving at velocity
near to velocity of light $c$), again giving $h_{\theta }$ the eigenvalues $%
\pm 1$. The evidence for helicity ($\pm 1$) of massless Dirac electron is
experimentally confirmed at Dirac point, such as in graphene \cite{A. H.
Castro Neto, A. K. GEIM}.

The intrinsic spin of electron is associated with the circular flow of
energy-momentum ($p_{\theta }^{o},\vec{p}_{\theta }$) of TWEM model. The
spin angular momentum $S$ is linked to the energy and angular frequency of
the circularly polarized wavepacket as follows \cite{Hans}

\begin{equation*}
S\equiv u_{em}/\omega ^{\prime }.
\end{equation*}%
With $p_{\theta }$ circulating in the $xy$-plane, it is straightforward to
show by using $u_{em}=\hslash \omega _{o}$\ and $\omega ^{\prime }=2\omega
_{o}$ that the $z-$component of spin is given by%
\begin{equation}
S_{z}=\pm \frac{1}{2}\hslash .  \label{spin}
\end{equation}%
where $\pm $\ stand for ccw and cw rotations, respectively. Alternatively,
with the geometry of Fig. (\ref{1}), we can determine the spin using the
internal orbital angular momentum generated due to circular flow of
energy-momentum%
\begin{equation*}
\vec{L}_{int}=\vec{r}\times (\vec{E}_{r}\mathbf{\times }\vec{H}_{z}\mathbf{)=%
}\vec{r}\mathbf{\times }\vec{p}_{\theta }=\left\vert \vec{r}\right\vert
\left\vert \vec{p}_{\theta }\right\vert \hat{\theta}.
\end{equation*}%
Using all the known values, this leads to $L_{int}=\frac{1}{2}\hslash $. For
free massless spinors, the spin eigenstates can also be taken as the
eigenstates of helicity with eigenvalues $\mathbf{\pm }1$. Correspondingly,
the $z$-components of spin have eigenvalues $\mathbf{\pm }1/2$, which
results in Eq. (\ref{spin}). Similar results for spin emerging from Ztg,
have been obtained in \cite{Bruno}. Also a topological structure of electron
has been hypothesized in \cite{Qiu Hong} and\textbf{\ }coinciding results
for spin have been obtained for postulated Hubius Helix model of electron.

\section{Electron in motion}

To see the behavior of electron wavefunction in motion, we apply the Lorentz
boost to Weyl spinors. The massless Weyl spinors seem to exist on the light
cone, moving at the speed of light. Furthermore, the velocity derivative in
proper time is zero, i.e. $\partial v/\partial \tau =0$ at the speed of
light. Applying boost to an object, already moving with speed of light is
pointless. However in the TWEM\ electron energy-momentum spinor(s) is
circulating around the origin (that is around the center of energy-momentum)
at the speed of light, Eqs. (\ref{8}, \ref{9}). Then, we can apply boost to
center of energy-momentum frame, which we consider, initially at rest. The
general Lorentz boost for spinors is given by \cite{Andrew}
\begin{equation}
S(\Lambda )=I\cosh \frac{\rho }{2}-\vec{\sigma}\cdot \hat{n}\sinh \frac{\rho
}{2},  \label{15}
\end{equation}%
where $\rho $ is rapidity such that $\tanh \rho =\beta $ and $\hat{n}$ is a
unit vector in the direction of motion. Using the relations $\cosh \frac{%
\rho }{2}=\left[ \frac{1}{2}(1+\gamma )\right] ^{\frac{1}{2}}$ and $\sinh
\frac{\rho }{2}=\left[ \frac{1}{2}(\gamma -1)\right] ^{\frac{1}{2}}$ along
with the relativistic relations of energy $E=\gamma m$ and momentum $\vec{p}%
=\gamma m\vec{v}=E$ $\vec{v}$, the Lorentz boost of Eq. (\ref{15}) in terms
of energy-momentum turns out to be
\begin{equation}
S(\Lambda )=\sqrt{\frac{E+m}{2m}}\left[
\begin{array}{cc}
I+\frac{\vec{\sigma}\mathbf{\cdot }\vec{p}}{E+m} & 0 \\
0 & I-\frac{\vec{\sigma}\mathbf{\cdot }\vec{p}}{E+m}%
\end{array}%
\right] .  \label{16}
\end{equation}%
Then we can write spinors $\psi _{R}(\vec{p})$ and $\psi _{L}(\vec{p})$ in
the moving frame as%
\begin{equation}
\left[
\begin{array}{c}
\psi _{R}(\vec{p}) \\
\psi _{L}(\vec{p})%
\end{array}%
\right] =S(\Lambda )\left[
\begin{array}{c}
\psi _{R}(0) \\
\psi _{L}(0)%
\end{array}%
\right] ,  \label{17b}
\end{equation}%
where $\psi _{R}(0)$ and $\psi _{L}(0)$ are spinors in the rest frame.\ For $%
\psi _{R}(p)$ and $\psi _{L}(p)$, we get%
\begin{equation*}
\left[
\begin{array}{c}
\psi _{R}(\vec{p}) \\
\psi _{L}(\vec{p})%
\end{array}%
\right] =\left[
\begin{array}{c}
\sqrt{\frac{E+m}{2m}}\left[ I+\frac{\vec{\sigma}\mathbf{\cdot }\vec{p}}{E+m}%
\right] \\
\sqrt{\frac{E+m}{2m}}\left[ I-\frac{\vec{\sigma}\mathbf{\cdot }\vec{p}}{E+m}%
\right]%
\end{array}%
\right] \left[
\begin{array}{c}
\psi _{R}(0) \\
\psi _{L}(0)%
\end{array}%
\right] .
\end{equation*}%
Assuming $\psi _{L}(0)=\psi _{R}(0)=\psi (0)$, we obtain massive Dirac
equation%
\begin{equation}
\left[
\begin{array}{c}
E-\vec{p}_{\theta }\cdot \vec{\sigma} \\
E+\vec{p}_{\theta }\cdot \vec{\sigma}%
\end{array}%
\right] \left[
\begin{array}{c}
\vec{\psi}_{R} \\
\vec{\psi}_{L}%
\end{array}%
\right] =\left[
\begin{array}{c}
m\vec{\psi}_{L} \\
m\vec{\psi}_{R}%
\end{array}%
\right] .  \label{17c}
\end{equation}%
In the moving frame the left and the right spinors are now mixed through the
mass term. Alternatively, we can mix up the left and the right spinors with
an appropriate weightage and can obtain the mass property of the massless
spinning (momentum) waves.

Now, for four component spinor $\vec{\psi}(\vec{p})\equiv \psi ^{\alpha }(%
\vec{p})$, we can write the massive Dirac electron equation in the momentum
space%
\begin{equation}
(\gamma ^{\mu }p_{\mu }-m)\vec{\psi}(\vec{p})=0.  \label{18}
\end{equation}%
In Minkowski space, replacing $p_{\mu }\longrightarrow i\partial /\partial
x^{\mu }=\partial _{\mu }$, $\mu =0,1,2,3$, the Dirac equation takes the
form
\begin{equation}
(i\gamma ^{\mu }\partial _{\mu }-m)\vec{\psi}(x)=0.  \label{19}
\end{equation}

Once the relation between energy-momentum of spinning EM fields with Dirac
equation is established,we can associate all attributes of Dirac theory to
the TWEM\ model of electron\textbf{.} The Dirac fields describing the model
are found on the divergence free symmetric energy-momentum tensor $\Theta
^{\alpha \beta }$of the circulating EM wave which obeys conservation laws $%
\partial _{\alpha }\Theta ^{\alpha 0}=\partial _{\alpha }\Theta ^{\alpha
i}=0 $ \cite{M.M.Asif}. There are number of ways to determine spin from
Dirac equations, however, we will follow \cite{Hans} to show it for circular
flow of energy-momentum of Dirac field. Using momentum density in Dirac
fields, the angular momentum $\vec{J}$ of electron (wave packet) becomes%
\begin{equation}
\vec{J}=\frac{\hslash }{2i}\int \vec{x}\times \lbrack \psi ^{\dag }\vec{%
\nabla}\psi -(\vec{\nabla}\psi ^{\dag })\psi ]d^{3}x+\frac{\hslash }{2}\int
(\psi ^{\dag }\vec{\Sigma}\psi )d^{3}x,  \label{19a}
\end{equation}%
where
\begin{equation*}
\vec{\Sigma}=\left(
\begin{array}{cc}
\vec{\sigma} & 0 \\
0 & \vec{\sigma}%
\end{array}%
\right) ,
\end{equation*}%
are Dirac spin matrices. In Eq. (\ref{19a}), the first part
represents orbital angular momentum and the second part represents
the spin. From the spin part of Eq. (\ref{19a}), the Dirac spin operator is obtained
as

\begin{equation*}
\vec{S}\mathbf{=}\frac{\hslash }{2}\vec{\Sigma}.
\end{equation*}%
The helicity operator $\hat{h}$ of Dirac electron is projection of spin on
linear momentum of electron, then

\begin{equation*}
\hat{h}=\frac{\vec{\Sigma}\cdot \vec{p}}{2\left\vert \vec{p}\right\vert }.
\end{equation*}

Alexander Burinskii \cite{Burinskii}\ has obtained similar results by
considering Kerr singular ring (a closed gravitational string) for the
electron model. He treats electron as electromagnetic excitation of the
ring, forming, travelling wave along a string as the Kerr-Newman solution of
Einstein-Maxwell's theory. The ordinary Dirac theory has been obtained from
the solutions of the massless underlying theory, which regards Ztg as a
corpuscular analogue of the travelling waves. The author attempts to
establish relation between gravity and QM leading to Quantum Gravity. The
Kerr singular ring model and TWEM\ model are vastly different in terms of
their modeling background theories. The Kerr singular ring model is based on
Kerr-Newman gravitational singular string whereas TWEM model is based on
linear Maxwell's theory and quantization.

In TWEM\ electron model, the dynamism (due to circular motion) of
energy-momentum gives rise to Ztg and associated properties, such as
electron spin and helicity etc. This model pictures, the internal structure
of electron (and positron) and interprets its underlying physical mechanism.
In perspective of the above deliberations, we may infer that TWEM electron
exhibits symmetry group $U(1)$ at traveling electromagnetic wave level \cite%
{M.M.Asif}. The $2D$ (circulating momentum) object belongs to symmetry group
$SU(2)$ in the rest frame as a massless Dirac spinor. Finally under boost
and/or rotation, $SL(2,C)$ symmetry group is pertinent to spinors leading to
massive Dirac theory. Hence electron is a complex object and exhibits blend
of different symmetry groups at different levels of structure and with
reference to frame of observation.

\section{Conclusion}

In summary, we have used the TWEM model to explore the internal structure of
electron. The circular traveling EM wave of the model carries, synchronous
rotating energy-momentum ($u,\vec{p}_{\theta }$), in a circle of radius of
half the reduced Compton wavelength of electron. It is demonstrated that in
TWEM model of electron, the Zitterbewegung and spin arise naturally as a
result of circular motion of the energy-momentum of EM wave. Furthermore,
the spinning energy-momentum wave function forms four component Dirac spinor
like object corresponding to the four possible circular polarized like EM
waves. For the four component energy-momentum spinor, the massless and
massive Dirac equations follow directly in the rest frame and in the boosted
frame, respectively. Alternatively, the four component spinor of the TWEM
model is a solution to the Dirac equation. In this model, we show that the
Dirac spinors are not just mathematical objects but are real objects with a
physical mechanism(s) in the background. The TWEM electron model is blend of
symmetry groups $U(1)$, $SU(2)$ and $SL(2,C)$, that is, different symmetries
are relevant at different levels of internal structure and with reference to
frame of observation.

\section{Acknowledgement}

M.M. Asif pays thanks to Robert L. Shuler Jr., NASA Johnson Space Center,
USA for his constructive suggestions regarding this work. The valuable
discussions with him, provided insight that greatly helped in carrying out
the present work.


\begin{thebibliography}{99}
\bibitem{Jansen} M. Jansen, and M. Mecklenburg, \textit{Electromagnetic
Models of the Electron}

\textit{and the Transition from Classical to Relativistic Mechanics}, Max
Planck

Institute for the History of Science, Berlin, (2004).

\bibitem{Barut} A.O. Barut and N. Zanghi, Phys. Rev. Lett. \textbf{52}, 2009
(1984).

\bibitem{Dirac2} P. Dirac, Proc. Roy. Soc. Lond. \textbf{209}, 291 (1951).

\bibitem{Dirac3} P. Dirac, Proc. Roy. Soc. Lond. \textbf{212}, 330 (1952).

\bibitem{Burinskii} A. Burinskii, J. Phys.: Conf. Ser. \textbf{361}, 012032
(2012).

\bibitem{Qiu Hong} Q.H. Hu, Physics Essays \textbf{17}, 442 (2004).

\bibitem{Dirac 1} P. Dirac, Proc. Roy. Soc. Lond. \textbf{117}, 610
(1928).

\bibitem{Dirac 4} P. Dirac, Proc. Roy. Soc. Lond. A \textbf{268}, 57 (1962).

\bibitem{Foldy L. L} L.L. Foldy and S.A. Wouthuysen, Phys. Rev. \textbf{78},
29 (1950).

\bibitem{Franz Schwabl} F. Schwabl, \textit{Advanced Quantum Mechanics}, 3rd
Ed. Springer-Verlag (2005).

\bibitem{E. Schrodinger} E. Schr\"{o}dinger, Sitzungsber. Preuss. Akad.
Wiss. Phys. Math. Kl. \textbf{24}, 418 (1930).

\bibitem{Barut and A. J} A.O. Barut and A.J. Bracken, Phys. Rev. D
\textbf{23} 2454 (1981).

\bibitem{Burinskii 1} A. Burinskii, J. Phys.: Conf. Ser. \textbf{343},
012019 (2012).

\bibitem{Burinskii 2} A. Burinskii, Grav. Cosmol. \textbf{14}, 109 (2008).

\bibitem{Atiyah} M.F. Atiyah, G. Franchetti and B.J. Schroers, JHEP. \textbf{%
02}, 062 (2015).

\bibitem{M.M.Asif} M.M. Asif, Phy. Essays \textbf{27}, 146 (2014).

\bibitem{Catillon} P. Catillon et al., Found. Phys. \textbf{38}, 659 (2008).

\bibitem{Gouan} M. Gouan\'{e}re et al., Annales de la Fondation Louis de
Broglie \textbf{30}, 109 (2005).

\bibitem{David Hestenes 1} D. Hestenes, Found. Phys. \textbf{40}, 1 (2010).

\bibitem{David Hestenes 2} D. Hestenes, Found. Phys. \textbf{20}, 1213 (1990).

\bibitem{Sasabe} S. Sasabe, J. Mod. Phys.\textbf{\ 5}, 534 (2014).

\bibitem{E. Romera} E. Romera, Phys. Rev. A \textbf{84}, 052102 (2011).

\bibitem{G. Cavalleri} G. Cavalleri, Lettere Al Nuovo Cimento Series \textbf{%
2}, 285 (1985).

\bibitem{David Hestenes 3} D. Hestenes, Found. Phys. \textbf{15},
63, (1983).

\bibitem{D. Hestenes} D. Hestenes, \textit{The Nature of Time Essay Contest}
(Foundational Questions

Institute, Decatur, GA, 2008).

\bibitem{K. Huang} K. Huang, Am. J. Phys. \textbf{20}, 479 (1952).

\bibitem{Zhi-Yong} Z.Y. Wang and A. Zhang, New interpretation to
Zitterbewegung, arXiv:hep-ph/0110079v1.

\bibitem{P.A.M. Dirac 1} P. Dirac, \textit{The Principles of Quantum
Mechanics}, Oxford University Press (1958).

\bibitem{R. Gerritsma} R. Gerritsma, G. Kirchmair, F. Z\"{a}hringer, E.
Solano, R. Blatt and C.F. Roos, Nature \textbf{463}, 68 (2010).

\bibitem{J. Y. Vaishnav} J.Y. Vaishnav and C.W. Clark, Phys. Rev. Lett.
\textbf{100}, 153002 (2008).

\bibitem{H. Torres-Silva} H.T. Silva, Ingeniare. Revista chilena de ingenier%
\'{\i}a, \textbf{16}, 72 (2008).

\bibitem{GIOVANNI} G. Salisi, Int. J. Mod. Phys. A \textbf{17}, 347 (2002).

\bibitem{Bruno} B.F. Rizzuti, E.M.C. Abreu, and P.V. Alves, Phys. Rev. D
\textbf{90}, 027502 (2014).

\bibitem{A. H. Castro Neto} A.H. Castro Neto, F. Guinea, N.M.R. Peres, K.S.
Novoselov and A.K. Geim, Rev. Mod. Phys. \textbf{81}, 109 (2009).

\bibitem{A. K. GEIM} A.K. Geim and K.S. Novoselov, Nature Materials \textbf{6}, 183 (2007).

\bibitem{Hans} H.C. Ohanian Am. J. Phys. \textbf{54}, 500 (1986).

\bibitem{Andrew} A.M. Steane, An introduction to spinors,
arXiv:1312.3824v1.
\end{thebibliography}
\end{document}